\documentclass{pasj00}

\begin{document}

\SetRunningHead{T. Kato, R. Ishioka, and M. Uemura}{Dramatic Changes in V503 Cygni}

\Received{}
\Accepted{}

\title{Dramatic Changes in the Outburst Properties in V503 Cygni}

\author{Taichi \textsc{Kato}, Ryoko \textsc{Ishioka}, and Makoto \textsc{Uemura}}
\affil{Department of Astronomy, Kyoto University,
       Sakyo-ku, Kyoto 606-8502}
\email{tkato@kusastro.kyoto-u.ac.jp}



\maketitle

\begin{abstract}
   We examined the VSNET light curve of the unusual SU UMa-type dwarf nova
V503 Cyg which is known to show a short (89 d) supercycle length and
exceptionally small (a few) normal outbursts within a supercycle.
In 1999--2000, V503 Cyg displayed frequent normal outbursts with
typical recurrence times of 7--9 d.  The behavior during this period
is characteristic to an usual SU UMa-type dwarf nova with a short supercycle
length.  On the other hand, V503 Cyg showed very infrequent normal
outbursts in 2001--2002.  Some of the superoutbursts during this period
were observed shorter than usual.  The remarkable alternations of
the outbursting states in V503 Cyg support the presence of mechanisms of
suppressing normal outbursts and premature quenching superoutbursts,
which have been proposed to explain some unusual SU UMa-type outbursts.
The observed temporal variability of the suppressing/quenching mechanisms
in the same object suggests that these mechanisms are not primarily
governed by a fixed system parameter but more reflect state changes in the
accretion disk.
\end{abstract}

\section{Introduction}

   ER UMa stars are a still enigmatic small subgroup of SU UMa-type
dwarf novae [for a review of dwarf novae, see \citet{osa96review}],
which have extremely short supercycle lengths
($T_s$ the interval between successive superoutbursts) of
19--50 d [for a review, see \citet{kat99erumareview}] and
regular occurrence of superoutbursts.
Only five definite members have been recognized up to now: ER UMa
(\cite{kat95eruma}, \cite{rob95eruma}, \cite{mis95PGCV});
V1159 Ori (\cite{nog95v1159ori}, \cite{pat95v1159ori});
RZ LMi (\cite{rob95eruma}, \cite{nog95rzlmi});
DI UMa (\cite{kat96diuma}); and IX Dra (\cite{ish01ixdra}).
Some helium-transferring cataclysmic variables
have become recognized as ``helium counterparts" of ER UMa stars
[CR Boo: \citet{kat00crboo}; V803 Cen: \citet{kat00v803cen},
\citet{kat01v803cen}].  From a theoretical side, ER UMa stars have been
understood as a smooth extension of normal SU UMa-type dwarf nova
toward higher mass-transfer rates ($\dot{M}$) \citep{osa95eruma}.
The exact origin of such a high mass-transfer rate is still a mystery.
Even considering a higher mass-transfer rate, the shortest period systems
(RZ LMi and DI UMa) are difficult to explain without a special mechanism
of prematurely quenching a superoutburst \citep{osa95rzlmi}.

   In recent years, there have been an alternative attempt to explain
the ER UMa-type phenomenon.  \citet{hel01eruma} tried to explain the
ER UMa-type phenomenon by considering a decoupling between the thermal
and tidal instabilities [see \citet{osa89suuma} for details of the
thermal-tidal instability model] under extremely small binary mass-ratio
($q$=$M_2$/$M_1$) conditions.  \citet{hel01eruma} speculated that
repeated post-superoutburst rebrightenings\footnote{
  These phenomena are sometimes referred to as {\it echo outbursts},
but we avoid using this terminology because this idea was first proposed
to describe the ``glitches" or ``reflares" in soft X-ray transients (SXTs)
\citep{aug93SXTecho}.  In SXTs, hard-soft transition is considered to
be more responsible for the initially claimed phenomenon
\citep{min96SXTtransition}, which is clearly different from dwarf nova-type
rebrightenings.
} in WZ Sge-type dwarf novae (hereafter WZ Sge stars) or
large-amplitude SU UMa-type dwarf novae
(e.g. \cite{kuu00wzsgeSXT}; see \citet{kat01hvvir} for a recent
observational review of WZ Sge-type stars).  \citet{bua01DNoutburst}
tried to explain ER UMa-type phenomenon by (rather arbitrary)
introducing an inner truncation of the accretion disk and irradiation
on the secondary star on a numerical model developed by
\citet{ham00DNirradiation}.  \citet{bua02suumamodel} further tried to
explain the unification idea by \citet{hel01eruma} using the same scheme
as in \citet{bua01DNoutburst}.  Although the results partly reproduced
the characteristics of ER UMa stars and WZ Sge stars,
they failed to quantitatively reproduce the light curves of these
dwarf novae.

   From the observational side, the existence of a gap between distributions
of ER UMa stars and ``usual" SU UMa-type dwarf novae has been a challenge.
The shortest known $T_s$ in usual SU UMa-type dwarf novae had been
130 d (YZ Cnc, see also Table 1 in \cite{nog97sxlmi}) at the time of the
initial proposition of ER UMa stars.  Although further works have slightly
shortened this minimum $T_s$ [SS UMi: 84.7 d, \citet{kat00ssumi};
BF Ara: 83.4 d, \citet{kat01bfara}], there still remains a undisputed gap.
In addition to these usual SU UMa-type dwarf novae with the shortest
$T_s$'s, there exists a seemingly different population of SU UMa-type
dwarf novae with short $T_s$'s, but with infrequent normal outbursts.
V503 Cyg [$T_s$ = 89 d, only a few normal outbursts in a supercycle
\citep{har95v503cyg}] and CI UMa [$T_s \sim$ 140 d, infrequent normal
outbursts \citep{nog97ciuma}; $T_s$ variable? \citep{kat02v344lyr}] are the
best-known examples.  The relation, however, between these objects and
ER UMa stars (and short $T_s$ usual SU UMa-type dwarf novae)
are unknown.  In most recent years, some instances of strong
$T_s$ variations have been reported in ER UMa stars
(\cite{fri99diuma}, \cite{kat01v1159ori}).  In this letter,
we report on the dramatic changes in the outburst properties in V503 Cyg.

\section{Observation and Analysis}

   We examined the observations of V503 Cyg posted to VSNET
Collaboration,\footnote{
$\langle$http://www.kusastro.kyoto-u.ac.jp/vsnet/$\rangle$.}
and found an appreciable change of the outburst properties.
The observations used $V$-band comparison stars, and typical errors
of individual estimates are smaller than 0.3 mag, which will not affect
the following discussion.  The object has been well sampled by many
observers around the world except for periods of solar conjunctions.

\begin{figure}
  \begin{center}
    \FigureFile(88mm,130mm){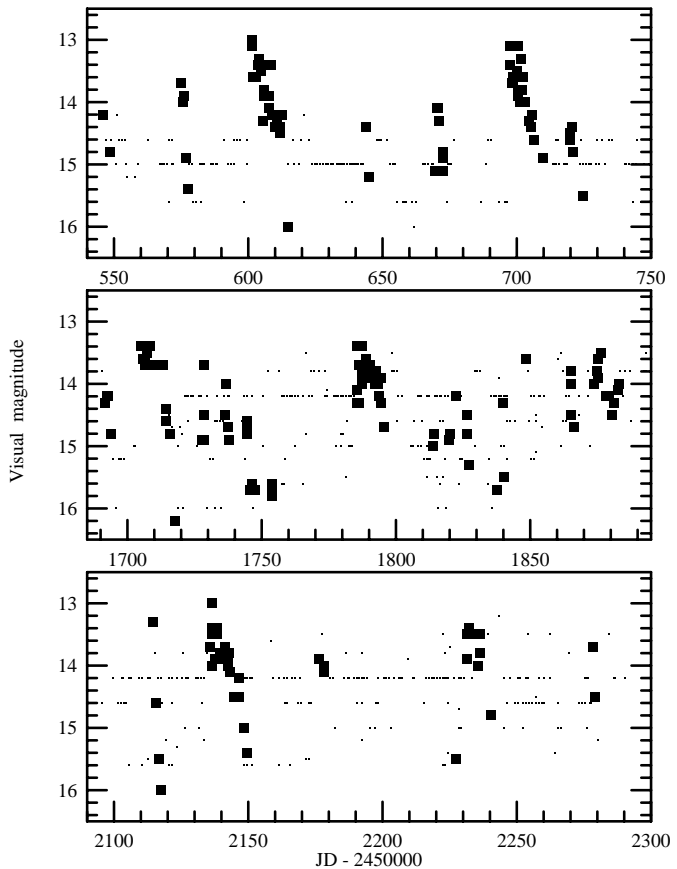}
  \end{center}
  \caption{visual light curve of V503 Cyg constructed from the
  observations reported to the VSNET Collaboration.  Large and small dot
  represent positive and negative (upper limit) observations, respectively.
  (Upper) ``Canonically" outbursting state in 1997.  Typically two
  normal outbursts are present within a supercycle (see also
  \cite{har95v503cyg}).  (Middle) State with increased number of normal
  outbursts in 2000.  The minimum cycle lengths of normal outbursts
  were 7--9 d.  (Lower) State with decreased number of normal outbursts.
  Note the short duration of the JD 2452231 superoutburst.
  }
  \label{fig:lc}
\end{figure}

   Table \ref{tab:burst} lists the observed outbursts since 1997.
The outburst lengths listed in the table approximately correspond to
the durations above $V \sim$ 15.5.  Although occasional observational
gaps introduced an uncertainty of a few days, most of these superoutbursts
were well recorded.  Many of normal outbursts during the favorably observed
seasons were recorded, although some outbursts must have been missed.
In order to estimate the numbers of missed outbursts, we performed
Monte-Carlo simulations.  The fractions of missing simulated 1000 normal
outbursts (the maximum magnitude and the rate of decline have been adjusted
to those of actual outbursts) were 10\%, 18\% and 30\%
for the three representative epochs shown in figure \ref{fig:lc}.
These fractions of missing normal outbursts will not affect the discussion
given in section \ref{sec:dis}.

\begin{table*}
\caption{Outbursts of V503 Cyg since 1997.}\label{tab:burst}
\begin{center}
\begin{tabular}{cccccccc}
\hline\hline
JD start & Peak mag & Length & Type &
JD start & Peak mag & Length & Type \\
\hline
2450545 & 14.2 & 3 & normal    & 2451519 & 14.4 & 2 & normal \\
2450574 & 13.7 & 3 & normal    & 2451531 & 13.8 & 10 & super \\
2450601 & 13.1 & 14 & super    & 2451641 & 14.5 & 1$^*$ & normal \\
2450643 & 14.4 & 2 & normal    & 2451666 & 14.5 & 2 & normal \\
2450669 & 14.1 & 3 & normal    & 2451673 & 14.4 & 3 & normal \\
2450697 & 13.1 & 13 & super    & 2451691 & 14.2 & 3 & normal \\
2450719 & 14.5 & 2 & normal    & 2451704 & 13.4 & 12 & super \\
2450758 & 14.3 & 1$^*$ & normal & 2451727 & 13.7 & 2 & normal \\
2450776 & 13.8 & 12 & super    & 2451736 & 14.0 & 2 & normal \\
2450896 & 14.3 & 2 & normal    & 2451744 & 14.6 & 3 & normal \\
2450952 & 13.3 & 10 & super    & 2451753 & 15.6 & 1 & normal \\
2450989 & 14.4 & 3 & normal    & 2451785 & 13.4 & 11 & super \\
2451018 & 14.1 & 3 & normal    & 2451813 & 14.8 & 2 & normal \\
2451051 & 13.3 & $>$8 & super  & 2451819 & 14.8 & 2 & normal \\
2451072 & 14.6 & 2 & normal    & 2451826 & 14.5 & 2 & normal \\
2451124 & 13.4 & 12 & super    & 2451837 & 14.3 & 2 & normal \\
2451220 & 14.0 & 1$^*$ & normal & 2451848 & 13.6 & 1$^*$ & normal \\
2451237 & 14.0 & 1$^*$ & normal & 2451865 & 13.8 & 2 & normal \\
2451323 & 14.5 & 3 & normal    & 2451873 & 13.6 & 11 & super \\
2451337 & 14.4 & 3 & normal    & 2451964 & 14.7 & 1$^*$ & normal \\
2451348 & 14.3 & 1 & normal    & 2452014 & 14.1 & 1$^*$ & normal \\
2451359 & 14.4 & 1$^*$ & normal & 2452046 & 13.5 & 13 & super \\
2451372 & 13.6 & 12 & super    & 2452114 & 13.3 & 3 & normal \\
2451406 & 14.4 & 2 & normal    & 2452135 & 13.0 & 14 & super \\
2451428 & 14.1 & 3 & normal    & 2452176 & 13.9 & 3 & normal \\
2451435 & 14.2 & 2 & normal    & 2452231 & 13.4 & 10 & super \\
2451453 & 13.7 & 13 & super    & 2452278 & 13.7 & 2 & normal \\
2451479 & 14.4 & 1$^*$ & normal & 2452321 & 13.8 & $>$5 & super \\
2451485 & 14.4 & 2 & normal    & 2452397 & 13.8 & 19$^\dagger$ & super \\
2451495 & 14.4 & 2 & normal    & 2452466 & 13.7 & 3 & normal \\
\hline
 \multicolumn{8}{l}{$^*$ Single observation.} \\
 \multicolumn{8}{l}{$^\dagger$ Brightening at the end. May have been two separate outbursts.} \\
\end{tabular}
\end{center}
\end{table*}

\section{Discussion}\label{sec:dis}

   In the standard disk instability model, the recurrence time of
normal outbursts ($T_n$) is mainly governed by the diffusion process,
while $T_s$ represents the increasing rate of net angular momentum
in the accretion disk \citep{ich94cycle}.  If the quiescent viscosity
parameter has a fixed value between various SU UMa-type dwarf novae,
both $T_n$ and $T_s$ are unique functions of $\dot{M}$
\citep{ich94cycle}.  This relation has been observationally confirmed
in most of SU UMa-type stars \citep{war95suuma}.  V503 Cyg apparently
violates this relation in its low frequency of normal outbursts
(figure \ref{fig:lc}, upper panel),
and several other stars (V344 Lyr, SX LMi) have been proposed to
be analogous to V503 Cyg (\cite{kat01sxlmi}, \cite{kat02v344lyr}).
There must be an unknown suppression mechanism of normal outbursts
in these systems.

   In 1999--2000, V503 Cyg showed a very frequent occurrence of normal
outbursts (minimum $T_n \sim$ 7--9 d, figure \ref{fig:lc}, middle panel).
This $T_n$ is just what is expected
for a $T_s$ = 89 d usual SU UMa-type dwarf nova \citep{war95suuma}.
This fact indicates that the usually outbursting SU UMa-type state
and unusually outbursting (in the sense of low frequency of normal
outbursts) V503 Cyg-type state are interchangeable.  Since $T_s$
during this period was not appreciably different from the canonical
$T_s$ = 89 d, there should have not been an appreciable change in the
$\dot{M}$. The suppression mechanism of normal outbursts must have been
somehow ``unlocked" during this period.

   In 2001--2002, V503 Cyg showed another different aspect
(figure \ref{fig:lc}, lower panel).  During this period, the number of
normal outbursts in a supercycle dramatically
decreased to $\sim$1.  There is some hint of alternating occurrence of
a superoutburst and a normal outburst with a period of 40--80 d.
Such a sequence of outbursts is only known in rarely outbursting
SU UMa-type dwarf novae [cf. SW UMa, V844 Her cf. \citet{kat00v844her}
for a discussion], and is unprecedented in short $T_s$ systems.
During this period, some normal outbursts have comparable peak magnitudes
to those of superoutbursts.  Some of superoutbursts showed rather
short durations, which seems to be incompatible with a high $\dot{M}$
necessary to reproduce the short $T_s$ \citep{osa95eruma}.
These findings suggest that premature quenching of superoutbursts,
as proposed by \citet{osa95rzlmi} and \citet{hel01eruma}, indeed
occurred during this period, although V503 Cyg (orbital period = 
0.0757 d) is unlikely to have a small $q$ required in \citet{osa95rzlmi}
and \citet{hel01eruma}.  The overall light curve more or less resembles
that of CI UMa (\cite{kol79cpdraciuma}, \cite{nog97ciuma}).

   Although exact mechanisms have not been yet identified, the present
remarkable alternations between the outbursting states in V503 Cyg support
the presence of mechanisms of suppressing normal outbursts and
premature quenching superoutbursts.  The most important finding is
that the effects of these mechanisms are temporarily variable even in the
same object, and are not a fixed character of a certain system.
This finding suggests that the shortest $T_s$ usual SU UMa stars and
unusual V503 Cyg-like stars can represent different aspects of the same
system.  Among ER UMa stars,
DI UMa can be a similar system with systematic state changes
\citep{fri99diuma}.  The observed temporal variability of the
suppressing/quenching mechanisms in the same object suggests that these
mechanisms are not primarily governed by a fixed system parameter
[i.e. mass of the white dwarf \citep{bua01DNoutburst};
$q$ \citep{hel01eruma} etc.] but more reflect state changes in the
accretion disk.

\vskip 3mm

   We are grateful to many amateur observers for supplying their vital
visual and CCD estimates via VSNET.
This work is partly supported by a grant-in-aid (13640239, TK) from the
Japanese Ministry of Education, Culture, Sports, Science and Technology.
Part of this work is supported by a Research Fellowship of the
Japan Society for the Promotion of Science for Young Scientists (MU).

\end{document}